\documentstyle[12pt]{article}
\begin{document}

\vspace*{2cm}
\centerline{ \Large GRAMA}

\bigskip

\centerline{ {\it Mathematica}  program
for  analytical}
\centerline{calculations in ten-dimensional supergravity
\footnote{Research supported by the ISF grant MOY000 }}

\bigskip

\centerline{Version 1.0}

\vspace*{2cm}

\centerline{\it User's Manual. }

\vspace*{2cm}

\centerline{\large   N.A.Saulina, E.M.Terentiev and  M.V.Terentiev  }

\vspace*{1cm}

\newpage

\vspace*{4cm}
\hspace*{3.5cm} \parbox{8cm}{ \begin{description}
\item ~~~~~~~~~~~Contents
\item  1. Introduction
\item  2. Program. Overall picture
\item  3. Main physical objects
\item  4. Main operators
\item 5. Appendix. \ \  Main expansion formulas
\end{description}
}

\newpage

\section{ Introduction}

"GRAMA" is a MATHEMATICA  package for doing symbolic tensor
computations and complicated algebraic manipulations in
ten-dimentional $(D=10)$, simple $(N=1)$ supergravity.  The main
new ingredients  of this package inside the general MATHEMATICA
invironment  is the  computation of complicated products
of Dirac matrices and treatment of covariant derivatives:
spinorial and  vectorial.

Other {\it Mathematica} packages for high-energy physics and gravitation
theory include: \cite{L},\cite{B},\cite{JL},\cite{HY},\cite{KBD},
\cite{MBD},\cite{M},\cite{STZ},\cite{T1},\cite{T2}.

The space-time dimension is introduced by the variable  "dim".
Analogously the dimension of spinorial representation is
introduced by  the variable "sdim".  In our case of $"dim"=10$ and
we use 16-component description of spinors, i.e.  $"sdim"=16$.
 In principle with small
modifications GRAMA may be applied for doing calculations in
the four-dimensional supergravity ("dim"=4, "sdim" = 4).

\section{ Program.  Overall picture}

The program consists from the following sections \footnote{There are
additional subsections of the type A1b.1, A1b.2, etc. in the program itself,
which are not shown below in the GRAMA (contents)}:

\bigskip
\centerline{"GRAMA" (contents): }

\bigskip

\begin{description}
\item PART A
\indexspace
\item  SECTION A1 ~~~~ Function "delprod".
\subitem sec. A1a.  Definition of "delprod"
\subitem sec. A1b.  Summation over dummy indices
\indexspace
\item  SECTION A2 ~~~~Levi-Civita "eps" tensor.
\indexspace
\item  SECTION A3 ~~~~Spinorial delta-symbol "sdelta".
\indexspace
\item  SECTION A4 ~~~~Function "gamprod"
\subitem sec. A4a.  Definition of "gamprod"
\subitem sec. A4b.  Some properties of "gamprod"
\subitem sec. A4c. Tracing of "gamprod"
\indexspace
\item  SECTION A5 ~~~~Summation over dummy indices in the "gamprod".
\subitem sec. A5a. Useful explicit formulae
\subitem sec. A5b.  Product of  "gamprod" with an
\subitem ~~~~~~~~~~arbitrary antisymmetric  function
\subitem sec. A5c.  The most general summation formula
\indexspace
\item  SECTION A6 ~~~~ Standard form of expressions.
\subitem sec. A6a. New properties of  "NonCommutativeMultiply"
\subitem ~~~~~~~~~~Anticommutive properties of fermionic functions
\subitem sec. A6b.  Fermionic functions
\subitem sec. A6c. Operator "newexp" transforming
\subitem ~~~~~~~~~  expressions to the "standard" form

\indexspace
\item  SECTION A7 ~~~~Physical functions
\subitem sec. A7a. Spintorsion: "STor[\{a,b\},Up[x]]" .
\subitem sec. A7b. Torsion: "Tor[\{a,b,c\}]"
\subitem sec. A7c.  Components of curvature tensor
\subitem ~~~~~~~~ "Rim[\{a,b\},\{c,d\}]", "Ric[a,b]",
\subitem ~~~~~~~~  "SRim[a,\{b,c\},Lo[x]]", "Rss[{a,b},Lo[x],Lo[y]]".
\subitem sec. A7d. Auxiliary antisymmetric function "Asm"
\indexspace
\item  SECTION A8 ~~~~Derivatives
\subitem sec. A8a. Vectorial derivative "vder"
\subitem sec. A8b.  Spinorial derivative "sder"
\subitem sec. A8c. Explicit formulas for spinorial derivatives
\subitem sec. A8d. Commutation relations between "vder" and "sder"
\indexspace
\item  SECTION A9 ~~~~Auxiliary operators
\subitem sec. A9a. "MySymm", "MyAsymm"
\subitem sec. A9b. "SameTerms1", "SameTerms2"
\subsubitem A9b.1. "AllvecIndex", "AllspinIndex", "AllgaugeIndex"
\subsubitem A9b.2. "DummyIndex"
\subsubitem A9b.3. List of  spinorial indices
\subsubitem ~~~~~~~  List of vectorial indices
\subsubitem ~~~~~~~   List of gauge indices
\subsubitem A9b.4. "ReplaceIndex"
\subsubitem A9b.5.  Definition of "SameTerms1", "SameTerms2"
\indexspace
\item PART B
\indexspace
\item SECTION B1 ~~~   Expansion formula for "gamprod"
\indexspace
\item SECTION B2 Constraints on "STor". Function "SAux"
\subitem sec. B2a. Cosequences of the constraint $ \Gamma^{ab}T_{ab} =0 $.
\subitem sec. B2b. Contraction of dummy indices in products of
-"gamprod" and "STor".
Function "SAux".
\indexspace
\item SECTION B3 ~~~Matrix Elements: Spintorsion - Spintorsion.
\subitem ~~~~~~~~~ ( "meTT", "meAT", "meTA", "meAA")
\indexspace
\item PART C
\indexspace
\item SECTION C1~~~ Matter fields "Glu" and "SGlu".
\item SECTION C2~~~ Spinorial derivatives of matter fields.
\item SECTION C3~~~Matrix elements:
\subitem ~~~~~~~~~~~ Gluino-Spintorsion, Gluino-Gluino
\subitem ~~~~~~~~~~~  ("meGA", "meGT", "meGG").
\indexspace
\item PART D
\indexspace
\item IDENTITIES
\end{description}

\bigskip

The program consists from four files: "progr-a.ma",  "progr-b.ma",
 "progr-c.ma",  "progr-id.ma", which are the {\em Mathematica} scripts.
 The main text
(PART A, SECTIONS  A1-A9) is contained in the file "progr-a.ma", the PART
B, SECTIONS B1 - B3, are  contained in the file  "progr-b.ma",
PART C is contained in the file "progr-c.ma" and PART D is contained in the
file "progr-id.ma". The last script ("progr-id.ma") contains
 useful identities, which
help to simplify expressions, containing $\Gamma$-matrices,
 torsion
and curvature superfields.

  The order, in which the files are loaded, is crucial for optimal
performance of GRAMA. Loading all the scripts "a", "b", "c" and "id"
simultaneously, or indiscriminately, may increase the number of
computation steps by an order of magnitude. The time of calculation of
any specific expression is much reduced, if these files are loaded
in sequence, as described below.

   First, one loads the "progr-a.ma" script and then the file or expresion
which must be calculated (simplified). (If this file/expresion contains matter
degrees of freedom, than the script "progr-c.ma" must be loaded simultaneously
with "progr-a.ma").
The calculation of the input expression with the help of the "progr-a.ma"
(and, if necessary, "progr-c.ma") scripts is the {\it first step} of the
procedure.
After it is finished, one should
load the file "progr-b.ma" and perform manipulations with
 the output obtained at the
first step. Calculations with the help of the "progr-b.ma" script is the
 {\it  second step}  of the calculation procedure.

 Finally, as the {\it third step}, the output of the second step is
simplified with the help of various identities, contained in the
"progr-id.ma" script.  That provides
 the final result of the GRAMA calculation procedure.

At each  {\it step}  of this procedure one should simplify outputs using
 operators Expand, SameTerms1, SameTerms2, etc, which are described in
the standard {\em Mathematica} handbook and in the Section 4 of this document.

\section{ Main physical objects }

We define here the correspondence between functions which are
used  in physical formulas and  functions used in the Program.

We are using  flat tangent-space vectorial
indices, so it is not necessary to fix their  position  (upper
or lower) in the MATHEMATICA notations if the rules of
contactions for identical (dummy) indices are defined
independently.

 In the 16-dimensional formalizm upper and
lower spinorial indices are distinguished. We use the notation
"Up[x]", for upper spinorial index "x", and "Lo[y]" for lower
spinorial index "y". We do not use in the Program an explicit
form of matrices, an explicit summation over dummy indices, etc.
All operations are defined by their properties  as in
corresponding theoretical formulas. We use  physical notations
and sign conventions from \cite{STZ} (these notations
corresponds  to that  from \cite{T1}, \cite{T2} up to the change
of sign of the curvature tensor components).
  Functions which one considers in
GRAMA  are superfields on the mass-shell, so their zero-superspace
components and zero components of
their  spinorial derivatives are unambiguously connected with
 physical fields of D=10  supergravity.

The list of main physical objects is presented in the table. The left
column is the MATHEMATICA notation, the right column is the
corresponding physical notation.

\bigskip

$$ \begin{array}{|c|c|}
 \hline
  delprod[\{a1 , \ldots , an \}, \{b1 , \ldots, bn \}]  &
    \delta^{\, a_1
 \ldots a_n}_{[b_1 \ldots b_n]}   \\
 \hline
 gamprod[\{a1,\ldots ,an\}, \ldots, \{c1, \ldots, cm\},Lo[x],Lo[y]]  &
 (\Gamma_{a_1
 \ldots\, a_n} \ldots
\Gamma^{c_1 \ldots\, c_m})_{xy}     \\
 \hline
 gamprod[\{a1,\ldots ,an\}, \ldots, \{c1, \ldots, cm\} ,Up[x],Lo[y]]  &
{(\Gamma_{a_1
 \ldots \,a_n} \ldots \Gamma^{c_1 \ldots\, c_m}  )^x}_y  \\
\hline
 eps[\{a1,\ldots,an\}] & \epsilon_{a_1\ldots a_n} \\
\hline
 sdelta[Up[x],Lo[y]] & {\delta^x}_y \\
\hline
 STor[\{a,b\},Up[x]] & (T_{ab})^x \\
\hline
SAux[a,Lo[x]] &  (T_a)_x \, ; \ \  T_a = \Gamma^b \, T_{ab} \\
\hline
 Tor[\{a,b,c\}] & T_{abc} \\
\hline
Tor2[a,b] & T_{acd}{T_b}^{cd} \\
\hline
Tor^2  & T_{abc}T^{abc} \\
\hline
SGlu[J,Up[x]] & (\lambda^{J})^x  \\
\hline
 Glu[\{a,b\},J] & {{\cal F}_{ab}}^{J} \\
\hline
 Rim[\{a,b\},\{c,d\}] & {\cal R}_{abcd} \\
\hline
Ric[a,b] & {\cal R}_{ab} = {{\cal R}_{acb}}^c \\
\hline
 SRim[a,\{b,c\},Lo[x]] & {\cal R}_{xabc} \\
\hline
 vder[ f, a] & D_a\, f  \\
\hline
 sder[ f, Lo[x]] & D_x\, f    \\
\hline
 Dil  & \phi \\
\hline
 SDil[Lo[x]] & \chi_x \equiv D_x \phi \\
\hline
meTT[\{a1,\ldots,an\}, \{a,b\},\{c,d\}]
 & T_{ab}\, \Gamma_{a_1\ldots a_n}\,T_{cd} \\
\hline
meTA[\{a1,\ldots,an\},\{c,d\} ,b] & T_{cd} \,\Gamma_{a_1\ldots a_n}\, T_b \\
\hline
meGT[\{a1,\ldots,an\}, J,\{c,d\}], \ \ etc. &
 \lambda^J \,\Gamma_{a_1\ldots a_n}T_{cd},  \ \ etc. \\
\hline
Asm[\{a1,\ldots, an\}] & Asm^{a_1\ldots a_n} \\
\hline
\end{array} $$

NOTE!: 1) one must use for vectorial indices the variables
start  with  letters: a,b,c,\ldots,q; \  spinorial
indices  start  with  letters: r,s,\ldots,z\  and enter
as the arguments of functions "Up[\ ]" and "Lo[\ ]"; \
 Yang-Mills indices  start  with
letters: I,J,K;\  2) all objects  containing odd number of
spinorial indices are fermionic in nature, the product of any
pair of fermionic objects $X$ and $Y$ must be represented by the
NonCommutiveMultiply function in MATHEMATICA, \  i.e. $X**Y$.
 3) "Asm[\{ \ldots\}]"
is an auxilliary completely antisymmetric function, which is
used for contraction of antisymmetric  external indices
 in any expression.

Some   comments  are presented below which are related with
 lines of the table.

1)  "delprod" is the completely antisymmetric product of
$\delta$-symbols. (GRAMA, Sec. A1).  The example is:
 $$
delprod [\{a,b\}, \{c,d\}]  \equiv \delta_{[a,b]}^{\, c,d}
\equiv {1\over 2!}\, (\delta_a^c\, \delta_b^d -
\delta_b^c \, \delta_a^d),    \eqno(3.1)$$

 The lists, which are  arguments of
"delprod", must have equal lengths, not exeeding "dim".
The Sec. 1  includes also
 contraction rules of "delprod" with an arbitrary
tensor.

2)  "gamprod" is the product of arbitrary  number
of completely antisymmetric products of
$\gamma_a$ matrices. (GRAMA, Sec. A4). The example is:
$$ gamprod [\{a,b\}, \{c,d,e\},Lo[x],Lo[y]] \equiv \,
 (\Gamma_{ab}\, \Gamma_{cde})_{xy}  \eqno(3.2)$$
where:
$$ \Gamma_{ab} \equiv {1\over 2!} (\Gamma_{a}\, \Gamma_b-
\Gamma_{b}\, \Gamma_a )\,
\ \ \ \Gamma_{abc} ={1\over3!}(\Gamma_a\Gamma_b\Gamma_c \pm permutations)
 \eqno(3.3) $$.

There is  the difference between spinorial and vectorial arguments
of "gamprod". Vectorial indices are always put in braces ("Lists" in
MATHEMATICA notations), this
is not the case for spinorial indices. There is an obvious
restriction for the number of vectorial  indices in each list
$Length \{ \ldots \} \leq  dim. $ But the total number of lists
is not limited.

The function "gamprod" with odd total number of vectorial indices contains
Up-Up or Lo-Lo spinorial indices, but "gamprod" with even total number
of vectorial indices contains Up-Lo or Lo-Up spinorial indices.

Symmetry properties of "gamprod" in vectorial and spinorial indices
are defined in the GRAMA.

3) "eps"-Levi-Chevita epsilon tensor (GRAMA,  Sec. A2).
 The number of arguments in the function eps[\{ \ldots \}]
 must be equal to "dim".

4) "sdelta"-spinorial delta-symbol (GRAMA, sec. A3).

5) "STor" and "SAux" are the fermionic (spin)
 torsion-components $(T_{ab})^x $ and $\Gamma^bT_{ab}$ (GRAMA, sec. A7a).
 $T_{ab}$ is antisymmetric in vector indices
and subjected to the
constraint: $\Gamma^{ab}\, T_{ab} = 0 $ (GRAMA, SECTION B2).

6) "Tor" is the torsion-component $ T_{abc} $  (GRAMA, sec. A7b).
It is the completely antisymmetric tensor.

7) "SGlu" is the fermionic chiral gluino  field $(\lambda^{J})^x$
 (GRAMA, sec. C1).

8) "Glu" is the gluon field-strength  ${{\cal F}_{ab}}^J $
which is antisymmetric in vectorial indices (GRAMA, sec. C1).

9) "Rim" is the super-curvature tensor $R_{abcd}$  (GRAMA, sec. A7c).
It has  usual symmetry properties and is  subjected to the
constaints:
 $$ {R_{ab}}^{cd}\, \delta^{ab}_{cd} =
 {1\over3}\, T_{abc}^2 \eqno(3.4) $$.
 $$Asm^{abc}\, (R_{abcd} - D_{a}T_{bcd}
 - T_{jab}{T^j}_{cd})=0  \eqno(3.5)$$
 $$ Asm^{ab}{R^j}_{ab}j = 0 \eqno(3.6)$$
 $$ Asm^{abc}\,(D_{a}R_{bcij}+
  T_{ab} \, R_{c ij}) = 0 \eqno(3.7)$$
(Here and below we do not write spinorial indices explicitely in the
cases, where their position may be reconstructed unambiguously).

Function "Ric" is a Ricci tensor.
(see GRAMA, sec. A7c).

10) "SRim" is the  spinorial components of the corresponding
superspace curvature $R_{ABCD}$. (GRAMA, sec. A7c.). The
"SRim" is not independent function:

$$ R_{abc} = 2\,\Gamma_{[b}\, T_{c]a} + {3\over 2}
  \Gamma_{[ab}\,T_{c]} \eqno(3.8) $$

11) "vder" and "sder" are vec\-torial and spino\-rial deri\-va\-tives
(GRAMA,\- Sec. A8).
The action of "sder" is de\-fined according to the following rules.
If "sder" appears to the left from "vder" then the GRAMA put it to the
right using the commutation relations (see \cite{STZ} and GRAMA sec. A8d).
The example is:

$$ (D_x\, D_b - D_b\, D_x)\, (V_c)_y  =
{1\over 72} T^{ijk}\,(\Gamma_{ijk}\,\Gamma_b\,D)_x\,(V_c)_y -  $$
$$- (R_{bij})_x \,({1\over4}\Gamma^{ij}\,V_c)_y - (R_{bcd})_x\, (V^d)_y - $$
$$ - (\Gamma_b\lambda)_x \,(V_c)_y - (V_c)_y\, (\Gamma_b\lambda)_x
\eqno(3.9)  $$
Here it is supposed that an arbitrary field $(V_c)_y$ is in the
algebra of Yang-Mills group G.

The application of "sder" to the main physical superfields is
calculated in according to the rules:

$$ D\,T_{abc} = 3\,\Gamma_{[a} T_{bc]}+
 3\,\Gamma_{[ab} \,  T_{c]}  \eqno(3.10) $$

$$ D_x\,(T_{ab})^y = {({\hat O}_{ab})^y}_x \eqno(3.11) $$
where:
$${\hat O}_{ab} =-{1\over 36}\Gamma_{[a} \Gamma^{ijk}D_{b]}\, T_{ijk}
+ {1 \over 36 \cdot 72}\,
 \Gamma_{[a} \Gamma^{mnp}\Gamma_{b]}  \Gamma^{ijk}\,
T_{mnp}T_{ijk} -$$
 $$+ {1\over 72}\, \Gamma^m  \Gamma^{ijk}  T_{abm}T_{ijk}-
{1\over 4} R_{abij}\Gamma^{ij} $$

and:
$$ D_x\lambda^y =
  {1\over 4}{\cal F}_{ab}{(\Gamma^{ab})_x}^y  \eqno(3.12)$$

Furthemore:
$$ D\,{\cal F}_{ab}= 2\Gamma_{[a}\,
D_{b]}\,\lambda-
T_{abc}\,\Gamma^c\,\lambda
-{1\over36}\,
 T^{ijk}\Gamma_{ijk}\Gamma_{ab}\,\lambda \eqno(3.13)$$

$$ D R_{abij} = 2\, D_{[a}\, R_{b]ij} +
{1\over36}\, T^{mns}\, \Gamma_{mns}\Gamma_{[a}\,
 R_{b]ij} +$$
$$+ R_{dij}\, {T_{ab}}^d
- ({5\over6}\, T_{ijk}\Gamma^k +{1\over36}\, T_{mnp}{\Gamma^{mnp}}_{ij})\,
T_{ab}  \eqno(3.14)$$

12) "meTT", "meTA", "meAA", "meGA", "meGT", "meGG"
 are matrix elements (see GRAMA SECTIONS  B3 and C3.).

 The GRAMA (see sec. A7d.) contains also the auxilliary
function "Asm". This function is a
"quasiconstant" (its vectorial and spinorial derivative is
zero, see SEC. A9) and completely antisymmetric in their arguments.  It is
used in the Program at  intrmediate stages for
contraction of external indices. It helps to avoid the
 antisymmetrization procedure in  external indices.

\bigskip

\section{ Main operators }

{\it NOTE! , Only the information, related with the "SameTerms1" and
"SameTerms2" operators is necessary for User, who is not interested
 in the internal structure of
 the Program. This information is contained in the 4) subsection
of this Section }

Here we are discussing new operators, which are defined in GRAMA
in additional to the standard MATHEMATICA operators.

1) Operator "newexp" (GRAMA, sec.A6c) transforms expressions to
the "standard" form.  We use the notation "standard" for the
expresion, where the symbol " ** " stands inside each pair of
fermions,  meanwhile any such pair is connected with other
pairs, or, with single fermion, or with scalars by means of " *
".  For instance (here s,r are any scalars):

$$ \mbox{newexp}[(s*SGlu)**(r*STor)]=
s*r*SGlu**STor $$

$$ \mbox{newexp}[(s*SGlu)**(SGlu*r*STor**STor)]=$$
$$=s*r*SGlu**SGlu*STor**STor $$

We are forced to work with the "standard" form, since the symbol
" * " means the commutative multiplication and can't stand
inside the pair of fermionic functions. Fermions in each pair
are automatically transposed according to the anticommutative
law so that their spinorial indices are put in lexicographical
order (GRAMA, sec.A6a ).

2) Operator "vderSimplify" (GRAMA, sec.A8a) serves for pulling
quasiconstants "gamprod",  "eps", "delprod", etc. from
under the symbol "vder".
(By definition of a quasiconstant objects, their
derivatives are equal to zero). Example:

$$ \mbox{vderSimplify}[vder[gamprod[...]*Tor[{a,b,c}]]] =$$
$$=gamprod[...]*vder[Tor[{a,b,c}]] $$

 In some cases (for example in commutation
relations of spinorial and vectorial derivatives) it is
convenient to keep quasiconstants under the symbol of "vder" and
to pull them from under this symbol only at the end of
calculations  with the help of operator "vderSimplify".

3) Operators "MySymm" ("MyAsymm") (GRAMA, sec.A9a)  make
symmetrisation (antisymmetrisation) of any expression in any
indices. For example

$$ \mbox{MySymm}[expression, \ a,b]\ \ \mbox{or} \ \
\mbox{MyAsymm}[expression, \ a,b]  $$
produce the expression, symmetric or antisymmetric in indices a,b.

4) Operators "SameTerms1", "SameTerms2" (GRAMA, sec.A9b) serve to
simplify any expression.  They make identical those terms in an
expression, which are  equal after rearrangement and
redefinition of dummy indices.  They differ from each other, and
one  must use both of them (one after another) in order to
identify all equal terms. (The first operator studies an
expression identifying identical terms  from the left to the right,
 the second one - from the right to
the left).

5) Operator "CountAll" gives the number of all fermions in an expression
(GRAMA, sec.A6b).

6) Operator "AllvecIndex" gives the list of all vectorial indices
in an expression.
Operator "AllspinIndex" gives the list of all spinorial indices
in an expression.

Operator "DummyIndex" is applied to the list of indices and gives the list
of repeating indices.

All these operators are used as auxiliary in the "SameTerms1", "SameTerms2".

\bigskip
\centerline{\bf \large Appendix}
\bigskip
\centerline{\bf \ Main expansion formulas}

\bigskip

We are using here simultaneously the physical notations and
notations from GRAMA defined according to the Table from Sec. 3
1. Symplification of products of gamma-matrices is realized by
the successive application of expansion formula:

$$ \Gamma_{a_1 \ldots a_n}\, \Gamma^{b_1 \ldots b_m} =$$
$$ \sum_{k=0}^{min(m,n)}  \eta_k\, (-1)^{k(n+1)}\, k!\,
\left( \begin{array}{l}
n \\ k
\end{array}\right)
\left(\begin{array}{l}
m \\ k
\end{array} \right)
\delta_{[a_1 \ldots a_k}^{[b_1 \ldots b_k}\ {\Gamma_{a_{k+1}\ldots a_n]}}^
{b_{k+1} \ldots b_m]}       \eqno(A1)    $$
$Traces$ are calculated according to relations:

$$ tr\ (\Gamma_{b_1...b_k}\Gamma^{a_1...a_k}) =
16\, k!\, \eta_k\, \delta_{\,b_1...\,b_k}^{[a_1...a_k]} ,
 \ \ \ k\not= 5   $$
$$ tr\ (\Gamma_{b_1...b_n}\Gamma^{a_1...a_m}) =0,  \ \ \ n\not= 10-m
\eqno (A.2)   $$
$${(\Gamma_{a_1...a_5}\Gamma^{b_1...b_5})_\alpha}^\alpha =
16\cdot5!\,\delta_{a_1...a_5}^{b_1...b_5} -
16\,{\epsilon_{a_1...a_5}}^{b_1...b_5}  \eqno (A.3)                  $$
$$ {(\Gamma_{a_1...a_k}\Gamma^{b_1...b_{10-k}})_\alpha}^\alpha =
-16\,{\epsilon_{a_1...a_k}}^{b_1...b_{10-k}} \eqno (A.4)               $$
In the  preceeding relations the sign before the $\epsilon$-term in the
r.h.s. should be changed to $+$ for ${(\ )^\alpha}_\alpha$ position of
indices in the l.h.s.

Note! {\it The "Trace" operator is not used in the program.} The {\it Trace}
is calculated automatically, if  User write a matrix with equal
upper and lower indices (i.e a matrix of the type
gamprod[\ldots , Up[x], Lo[x] ] or gamprod[\ldots , Lo[x], Up[x] ] ).

In the process of calculations we are using the equation (A1) and other
similar equations
in the form where all the antisymmetrisators $[\ldots ] $ are
calculated explicitely with the help of the relation:

$$ f([a_1 \ldots a_k)\, g(a_{k+1}\ldots a_n]) =
 \sum_{j_k,j_{k-1},\ldots , j_1 =1}^{n}$$ $$ \, (-1)^{k(k+1)/2}\,
(-1)^{(j_1 + \ldots + j_k) }\  {k!(n-k)!\over n!}
\, f(a_{j_1} \ldots a_{j_k})\,
g(a_1 \ldots {\dot a}_{j_1} \ldots {\dot a}_{j_k}\ldots a_n)
    $$
$$j_k > j_{k-1} > \ldots > j_1\, ; \ \ \  n-k \geq k  \eqno(A5')   $$

$$ f([a_1 \ldots a_k)\, g(a_{k+1}\ldots a_n]) =
 \sum_{s_{n-k}, s_{n-k-1},\ldots , s_1 =1}^{n} (-1)^{k(n-k)}\,
(-1)^{(n-k)(n-k+1)/2}$$
 $$ \,
(-1)^{(s_1 + \ldots + s_{n-k}) }\  {k!(n-k)!\over n!}
\, f(a_1 \ldots {\dot a}_{s_1}\ldots {\dot a}_{s_{n-k}} \ldots a_n)\,
g(a_{s_1} \ldots  a_{s_{n-k}})
    $$

$$s_{n-k} > s_{n-k-1} > \ldots > s_1\, ; \ \ \  n-k \leq k  \eqno(A5'')   $$
where  arguments should be cut away from their places in the
cases, when they are dotted,  $f(\ldots)$ and $g(\ldots)$ are
any completely antisymmetric functions of their arguments ( In
the case of eq. (A1) these functions play the role of $gamprod$
and $delprod$). Eq's (A5') and (A5'') help
  to reduce the $n!$ terms in the
antisymmetrizator in the r.h. side of (A1) into the $n!\over k!
(n-k)!$ terms in the r.h. side of (A5). In the case $ n=2k $
eq's (A5'), (A5") give the same result.

All the formulas (A1)-(A5) (i.e. file "B")
 are applied $after$ the contraction of
all dummy indices is fulfilled (this succession helps to decrease the
 calculation time).

The contaction of indices in the $gamprod$'s is fulfilled with the help
of the relation:

$$ gamprod[\{a_1\ldots a_k b\},\{b d_1\ldots d_m\}] =$$
$$-\sum_{j=1}^k (-1)^{k-j}\ gamprod[\{a_1\ldots {\dot a}_j \ldots a_k\},
\{a_j d_1\ldots d_m\}] +     $$
$$+(d-m) \,gamprod[\{a \},\{ d \}]    \eqno(A6) $$
And in more general case:
$$gamprod[\{a_1\ldots a_k \nu\},\{b_1\ldots b_n\}, \{c_1\ldots c_m\},
\{ \nu d_1\ldots d_p\}] = $$
$$= - \sum_{j=1}^k (-1)^{k-j}\, gamprod[\{a_1\ldots {\dot a}_j \ldots a_k\},
\{b \},\{c \}, \{a_j d_1 \ldots d_p\}]  - $$
$$- 2\,(-1)^n \sum_{j=1}^n (-1)^{n-j} gamprod[\{a \},
\{b_1\ldots {\dot b}_j \ldots b_n\}, \{c \},
\{b_j d_1\ldots d_p\}]- $$
$$-2\,(-1)^{n+m}\sum_{j=1}^m (-1)^{m-j} gamprod[\{a \},\{b
\},
\{c_1\ldots {\dot c}_j\ldots c_m\},\{c_j d_1\ldots d_p\}] +$$
$$+ (-1)^{m+n} (d-p)\, gamprod[\{a \},\{b \},
\{c \},\{d \}]           \eqno(A7)   $$
where $ \{a \}= \{a_1 \ldots a_k\}, \ \{b\}=\{b_1 \ldots b_n\}$, etc.

More general formula, where $gamprod$ contains an arbitrary nuber of
arguments is a direct generalization of eq. (A7).

The contraction of dummy indices in the $delprod$ is realized by
the  relation:

$$ \delta^{[a_1\ldots a_p c_1\ldots c_k]}_{[a_1\ldots a_p
d_1\ldots d_k]} = {(dim -k)!\,k! \over (dim -k-p)!\,(k+p)!}
\delta^{[c_1\ldots c_k]}_{[d_1 \ldots d_k]}   \eqno(A8) $$

The action of $delprod$ is defined by the relation:

$$ \delta^{a_1\ldots a_ka_{k+1}\ldots a_n}_{b_1\ldots b_k c_{k+1}\ldots c_n }
f^{c_{k+1}\ldots c_nc_{n+1}\ldots c_p}
 =\delta_{\,b_1\ldots b_k}^{[a_1\ldots a_k}
f^{a_{k+1}\ldots a_n]\,c_{n+1}\ldots c_p} \eqno(A9) $$
where $f_{c_{k+1}\ldots c_p}$ is an arbitrary completely antisymmetric
tensor.
. The further simplification of the r.h.s. of (A9)
must be realized with the help of (A5).


\begin{thebibliography}{99}
\bibitem{L} John M.Lee: Ricci. A Mathematica package for doing
calculations in differential geometry.  Dept. of Mathematics, GN-50.
University of Washington, Seattle, WA 98195.
 Internet: lee@math.washington.edu.
\bibitem{B} J.Beringer: Physica. A Mathematica package for the symbolic
calculation of tree-order processes. Preprint Univ. of Bern, BUTP-92/7.
Inst. for Theor. Physics, Univ. of Bern, CH-3012.
\bibitem{JL} M.Jamin, M.E.Lautenbacher: Tracer. A Mathematica package for
$\gamma$-algebra in arbitrary dimensions. Preprint Tech. Univ. Munchen,
TUM-T31-20/91.
\bibitem{HY} A.Hsieh, E.Yehudai: HIP-Symbolic High Energy Calculations,
SLAC preprint, SLAC-PUB-5576 (T/E).
\bibitem{KBD} J.Kublbeck, M.Bohm, A.Denner: FeynArts. Comp. Phys. Comm.,
vol 60 (1990) 165.
\bibitem{MBD} R.Mertig, M.Bohm, A.Denner: FeynCalc. Comp. Phys. Comm.
vol 64 (1991) 345.
\bibitem{M} R.Mertig: Guide to FeinCalc 1.0
\bibitem{STZ}\ N.A.Saulina, M.V.Terentiev, K.N.Zyablyuk: ~~Phys. Lett. 1995
   (in press)
\bibitem{T1}\ M.V.Terentjev:  ~~Phys. Lett. {\bf B 325} (1994)  96.
\bibitem{T2}\ M.V.Terentjev:  ~~Intern. J. Mod. Phys. {\bf A 9} (1994)  239.

\end{thebibliography}
\end{document}